\documentclass[12pt]{article}
\usepackage{amsmath}
\usepackage{amssymb}
\begin{document}
\title{Phase Transition in Light-Front $\phi^4_{1+1}$}
 
\author{V.~T.~Kim\\
{\it St.  Petersburg Nuclear Physics Institute,}\\
{\it Gatchina 188300, Russia}\\
{\it E-mail: kim@pnpi.spb.ru}
\and G.~B.~Pivovarov\\
{\it Institute for Nuclear Research,}\\
{\it Moscow, 117312 Russia}\\
{\it E-mail: gbpivo@ms2.inr.ac.ru}
\and J.~P.~Vary\\
{\it Department of Physics and Astronomy,} \\
{\it Iowa State University, Ames, Iowa 50011, USA}\\
{\it E-mail: jvary@iastate.edu}
}
 
\date{October 14, 2003}
\maketitle
 
\abstract{We reproduce Chang's duality condition in a
regularized $\phi^4_{1+1}$ theory quantized on a light front.
The regularization involves higher derivatives in the
Lagrangian, renders the model finite in the ultraviolet,
and does not require introduction of a finite size of the system.
It is demonstrated that the light-front quantization is a natural way
to treat systems with higher derivatives.
The phase transition is related to the presence of tachyons
in the regularized theory.
Prospects for computing the critical
coupling in this formulation are briefly discussed.}
 
\section{Introduction}
 
Light-front quantization promises to become an alternative (to
lattice formulation) computational approach to
quantum field theories (for a review, see
\cite{Brodsky:1997de}). The common objection against it is
that it predicts triviality of the vacuum, and, therefore, has trouble
addressing the existence of phase transitions in field theories.
A good test case to study this issue is the $\phi^4_{1+1}$ model.
For this model, the presence of a second order phase transition is proved
rigorously \cite{Glimm:ng}, and demonstrated
\cite{Chang:1976ek} with a
duality relation derived between the couplings of the theories
with different signs of the mass squared term in the Lagrangian.
The duality condition
implies
the two theories with different signs on the mass squared term
are identical. The duality relation maps the strong coupling
limit of the theory with positive mass squared (i.e., with a single
minimum
of the potential) to the small coupling
limit of the theory with a negative mass squared (i.e., with two minima
of the potential). There is also a
lattice computation for the value of the critical coupling
\cite{Loinaz:1997az}.
 
If the light-front quantization is a viable scheme for nonperturbative
computations, it should reproduce
both the duality condition, and the numerical value of the critical
coupling from the lattice
computation. There are attempts in the literature within
light-front quantization to achieve this
(see \cite{Robertson:1992nj, Heinzl:1995xj, Harindranath:ex, Bender:yd}).
 
Lately, the prevailing view maintains that the zero mode in
light-front $\phi^4_{1+1}$ theory accounts
for the properties of the phase transition
via its non-linear constraint
(see, for example, \cite{Bender:yd}). In this  "discretized
light-cone quantization" or "DLCQ" approach, the system is
put in a finite box along a light-like direction.
According to one report
\cite{Heinzl:1995xj}, DLCQ is unable to reproduce the lattice
results for the critical coupling, and yields a
wrong value for the critical exponent.
An alternative DLCQ investigation
developing the framework of
\cite{Rozowsky:2000gy} is now under way, and has given
promising results \cite{DHMPV}).
 
Another objection is that the
treatment of the zero modes is far from clearcut in that they
are expressed in terms of the dynamical
modes with a classical solution, and, subsequently, these expressions
are treated as operators.

A more intuitive objection against the traditional
form of DLCQ is that it seems to be at odds with the
more conventional (equal-time and Euclidean lattice) approaches.
For example, in the lattice computation of $\phi^4_{1+1}$ theory,
the presence of the phase transition is closely related to the
ultraviolet divergence present in the
one divergent diagram in this model.
The same relation to the divergent tadpole
diagram shows up in the derivation of the duality relation due to Chang
\cite{Chang:1976ek}.

The relation of the ultraviolet divergence to the phase transition
seems to be a fundamental feature, which one
may expect to show up in any treatment. At present,
there is no obvious connection
between the zero mode DLCQ treatment of the phase
transition and the ultraviolet divergence of the model.
Furthermore, to our knowledge, there is no
derivation of Chang's duality relation in the DLCQ treatment.
 
Reproducing Chang's duality condition constitutes a
challenge for DLCQ because the ultraviolet divergence
is independent of the mass of the excitation appearing
in the spectrum at small coupling. The role of the mass in the
DLCQ ultraviolet counterterm is played by
the inverse size of the system along the
light-like direction. But the phase transition emerges only in the
limit of the infinite size of the system. We conclude that
DLCQ regularization
may not accomodate straightforwardly the phase transition.
 
An alternative approach is considered in \cite{Rozowsky:2000gy}, where
the zero mode is discarded and a state of a soliton anti-soliton
pair is constructed for
the model with the negative sign of the mass term. Here the role of
the infinite size limit in the emergence of the phase transition is
stressed, but the role of the ultraviolet divergence is unclear, and there
is
no connection made with Chang's duality relation.
 
An approach sustaining a contact with the equal-time formulation is
developed in \cite{Harindranath:ex}. In this paper, the light-front
treatment
of the phase transition is related
to the ultraviolet divergence, the divergence is regularized, and Chang's
dualtiy relation is reproduced in light-front quantization.
However, this approach differs from the DLCQ
formulation,  so one loses
contact with the computational simplicity of DLCQ.

In this paper, we present a novel approach to
light-front quantization of $\phi^4_{1+1}$.
In a way, it continues the attitude of \cite{Harindranath:ex}
that the regularization
of the ultraviolet divergence of the theory is a crucial ingredient.
In contrast to \cite{Harindranath:ex}, we use a nonperturbative
regularization introduced
on the level of the Lagrangian of the model. One option is to use
regularization with a
light-front lattice \cite{Mustaki:gi}. We choose not to follow
this option  here,
because it breaks conservation of light-front momentum.
To retain the conservation
of the light-front momentum seems to be crucial, since defining
the state space
in sectors of definite
light-front momentum is the foundation of the
computational capacity  of light front quantization.
Instead, we chose to
regularize the theory with higher derivatives and to
conserve light-front momentum.
 
The light-front formulation is a natural way to quantize
theories with higher derivatives because the
number of light-front time derivatives is two times smaller
than the corresponding number of the derivatives
in the conventional physical time. Therefore, light-front quantization
of a theory whose Lagrangian
is quadratic in the Laplacian of the field is still a
quantization of a theory whose equations of motion involve only up to
second
derivatives over the light-front time. The use of the light-front
quantization
to quantize the theories with higher derivatives is one of the key
suggestions
of this paper.
 
The quantization we present here leads to the presence of tachyons in the
spectrum. Their mass parameter goes to infinity when the higher derivative
terms are switched off (i.e., when the ultravilet regularization
is removed). For this reason, the tachyons decouple in the limit of
the regularization removed. Still, they influence the dynamics of the
real particles. In particular, if the tachyon dynamics
implies a phase transition, the vacuum will become a tachyon condensate
after the phase transition.
Among the tachyons, there are excitations with small
longitudinal momenta. In
this way, the present approach does not deny the importance of the
zero modes (i.e., the modes of zero light-front momentum) for the phase
transition, but helps to consider them in a new way. We stress that
there are no constraints in this light-front  Hamiltonian
treatment
of the regularized theory, and there are no nondynamical degrees of
freedom
in this treatment.
 
This paper formulates our approach, reproduces
Chang's duality condition,
and discusses the prospects for transforming  our approach
into a computational scheme. In the next section,
we discuss our choice of the
ultraviolet regularization. In the third section, we give the regularized
model the light-front Hamiltonian treatment, i.e., define the
canonincal variables, the light-front Hamiltonian, and the light-front
longitudinal momentum. In the fourth section, Chang's duality
condition is reproduced in a way that mimics closely the original
derivation
due to Chang. In the last section, we discuss what
is needed to transform our approach
into a computational scheme.
 
\section{The Ultraviolet Regularization}
 
The Lagrangian [= density] of the theory is
\begin{equation}
\label{lagrangian}
{\cal L} = \frac{1}{2}g^{\mu\nu}\phi_{,\mu}\phi_{,\nu} - V(\phi),
\end{equation}
where $g^{\mu\nu}$ is the Lorentz metric tensor (its only nonzero
components are $g^{+-}=g^{-+}=1$). Hereafter, the indexes after
a coma in a subscript on a field denote the partial derivatives,
e.g., $\phi_{,\mu}\equiv\partial_\mu\phi$. The potential of the model is
\begin{equation}
\label{potential}
V(\phi)=\frac{m^2}{2}\phi^2 + \frac{g}{4}\phi^4.
\end{equation}

Perturbatively, there is a single divergent diagram in this model.
This is the tadpole diagram that appears from a pairing of two fields
involved in the same $\phi^4$ vertex. Analytically, it is
\begin{equation}
\label{tadpole}
T=\int\frac{dk_+dk_-}{(2\pi)^2}\frac{i}{k^2 - m^2 + i\delta},
\end{equation}
where $k^2=2k_+k_-$ is the Lorentz invariant  momentum squared. The
tadpole
is divergent for two reasons. First, the propagator decays too
slowly
at infinite $k^2$. Second, the propagator is Lorentz invariant. To
understand
this, assume that a term $(k^2)^2/M^2$ is introduced as a regulator in
the denominator of (\ref{tadpole}). Go over to new integration variables,
$(k_+,k_-)\rightarrow (k^2, k_-)$. Now the integral over $k^2$ converges,
but the integral over $k_-$ is still divergent. The reason for this is the
infinite volume of the group of Lorentz transformations.
The extra divergence related to the infinite volume of the
Lorentz group generally appears if a regularization breaks
the conventional structure of the poles of Feynman integrands,
and prevents the use of the Euclidean formulation.
We conclude that the use of ultraviolet regularization
with higher derivatives requires breaking of Lorentz invariance.
\footnote{We thank Prof. A. Vainshtein for discussing this point.}
 
Our choice for the regularized Lagrangian [= density] is
\begin{equation}
\label{regularized}
{\cal L}_r = \frac{1}{2}g^{\mu\nu}\phi_{,\mu}\phi_{,\nu}+
              \frac{1}{2}\epsilon t^{\mu\nu}\phi_{,\mu}\phi_{,\nu}+
               \frac{1}{2M^2}(\square\phi)^2
              - V(\phi),
\end{equation}
where $\square\phi=g^{\mu\nu}\phi_{,\mu\nu}$. The regularization is
removed
in the limit $\epsilon\rightarrow 0, M\rightarrow\infty$. As we will see,
upon quantization, $M$ becomes the mass parameter of the tachyons. The
tensor $t^{\mu\nu}$ is a symmetric tensor whose nonzero components in
a Lorentz frame are $t^{++}=1, t^{--}=-1$. The role of the term with
the $t$-tensor is to break the Lorentz invariance. The signs of the
components of this tensor are chosen to have a nonnegative light-front
Hamiltonian (see below).
 
In the next section, we perform the light-front quantization of the
model with the Lagrangian (\ref{regularized}). This quantization leads
to the field propagator of the following form:
\begin{equation}
\label{propagator}
\tilde{G}_r(k) = \frac{i}{k^2-m^2+\epsilon k^2_t+(k^2)^2/M^2+i\delta},
\end{equation}
where $k^2_t\equiv t^{\mu\nu}k_\mu k_\nu$.
With this propagator, the tadpole
\[
T=\int\,\frac{dk_-dk_+}{(2\pi)^2}\,\tilde{G}_r(k)
\]
is finite. Feynman diagrams not involving the tadpole have finite limits
when the regularization is removed. These limits coincide with the
corresponding diagrams of the original theory with the Lagrangian
(\ref{lagrangian}). From this we conclude that at least perturbatively
the regularized theory reproduces the original theory at the
limit the regularization is
removed.
 
\section{Light-Front Quantization of the Regularized Theory}
 
Let us start the quantization procedure by writing
down the expression
for the density of the energy momentum tensor of the regularized theory.
The Lagrangian [= density] (\ref{regularized}) 
is expressed in terms of the field
and derivatives of the field. The first and the second derivatives
of the field are involved.
Noether's procedure yields
\begin{equation}
\label{neother}
(\theta_r)^\mu_\rho=\frac{\partial{\cal L}_r}{\partial\phi_{,\mu}}
\phi_{,\rho} + \frac{\partial{\cal L}_r}{\partial\phi_{,\mu\nu}}
\phi_{,\nu\rho}-\big(\partial_\nu
\frac{\partial{\cal L}_r}{\partial\phi_{,\mu\nu}}\big)\phi_{,\rho}-
{\cal L}_r\delta^\mu_\rho.
\end{equation}
An integral of this density gives the light-front component
of the total momentum of the system:
\begin{equation}
\label{pminus}
P_-=\int\,dx^-(\theta_r)^+_{-}=\int\,dx^-\big[(\phi_{,-})^2+
\big(\epsilon\dot{\phi}-
\frac{2}{M^2}\square\phi_{,-}\big)\phi_{,-}\big].
\end{equation}
The overdot above denotes the derivative over $x^+$, which is considered
as dynamical time, and the line $x^-$ taken at a fixed value of the
light-front time is the manifold where the initial
condition for the field is set.
Similarly, for the momentum component along the  plus direction, we obtain
\begin{equation}
\label{pplus}
P_+=\int\,dx^-(\theta_r)^+_{+}=\int\,dx^-
\big[\frac{\epsilon}{2}\big(\dot{\phi}^2+(\phi_-)^2\big) +
\frac{1}{2M^2}(\square\phi)^2 + V(\phi)\big].
\end{equation}
 
The canonical coordinates of the regularized model are the values
$\phi(x^-)$ (their dependence on $x^+$ describes the dynamics). The
conjugated momenta are obtained as the variational 
derivatives of the Lagrangian 
[= in $\dot{\phi}$]
(\ref{regularized}) over $\dot{\phi}$:
\begin{equation}
\label{momenta}
\pi=\phi_{,-} + \epsilon\dot{\phi}-\frac{2}{M^2}\square\phi_{,-}.
\end{equation}
 
The above components of the total momentum expressed in terms of the
canonical variables are
\begin{equation}
\label{pminuscanonocal}
P_-=\int\,dx^-\pi\,\partial_-\phi,
\end{equation}
\begin{equation}
\label{ppluscanonical}
P_+=\int\,dx^-\big[\frac{1}{2}\,(\pi-\phi_{,-})
\frac{1}{\epsilon-4\partial_-^2/M^2}(\pi-\phi_{,-}) +
\frac{\epsilon}{2}\,\phi_{,-}^2 + V(\phi)\big].
\end{equation}
 
The expression for $P_-$ coincides with the one for the momentum space
component in the equal-time quantization (with the natural replacement
of the integral over the space by the integral over the light-front).
We use this observation
to diagonalize the momentum $P_-$. It is achieved with
 the following decomposition of the field and the conjugated momentum:
\begin{equation}
\label{creationannihilation}
\phi(x^-)=\int\frac{dl}{\gamma_m(l)\sqrt{4\pi}}\big
[a_l\exp{(-ilx^-)}+a_l^\dagger\exp{(ilx^-)}\big],
\end{equation}
\begin{equation}
\label{momentumcreation}
\pi(x^-)=\int\frac{dl}{i\sqrt{4\pi}}\,\gamma_m(l)\,\big
[a_l\exp{(-ilx^-)}-a_l^\dagger\exp{(ilx^-)}\big].
\end{equation}
Here $l$ has the meaning of the longitudinal momentum of the excitation
(see below), and ranges from negative to positive infinity.
The $\gamma_m(l)$ is an arbitrary real even function of the longitudinal
momentum $l$ at this stage. We will specify it when we
diagonalize the free light-front Hamiltonian. Notice that the
quantization proceeds in complete analogy
with the equal-time quantization. As in the equal-time quantization,
it is possible to diagonalize the free Hamiltonian (the role of the
Hamiltonian is played in the light-front quantization by $P_+$) by fitting
the dependence of $\gamma_m(l)$ on $l$. As we will see in a moment, this
dependence will be different at different masses. This is why we
put the subscript $m$ on $\gamma_m(l)$. Turning back to $P_-$, in terms of
the
creation-annihilation operators introduced above, it is
\begin{equation}
\label{pminuscreation}
P_-=\int\,dl\,l a_l^\dagger a_l.
\end{equation}
We see that $a^\dagger_l$ creates the excitation whose momentum component
along the direction $x^-$ is $l$, and the negative values of $l$ are not
forbidden. At the same time, $P_+$ defined in (\ref{ppluscanonical})
is explicitly positive. Thus, the excitations with negative $l$ are
tachyons
(i.e., their mass squared, $M^2=2P_-P_+$, is negative).
 
Now let us diagonalize the free Hamiltonian. To this end,
put $V(\phi)=m^2\phi^2/2$ into (\ref{ppluscanonical}), substitute the
expansions of the field and canonical momentum over the
creation-annihilation
operators, and require that the coefficient of the combination
$(a_la_{-l}+a^\dagger_la^\dagger_{-l})$ vanishes. This yields
the expression for $\gamma_m(l)$:
\begin{equation}
\label{gamma}
\gamma_m(l)=\big[l^2+(m^2+\epsilon l^2)(\epsilon + 4 l^2/M^2)\big]^
{\frac{1}{4}}.
\end{equation}
Again notice that the equal-time quantization formally differs only by
this expression (in the equal-time, $\gamma_m(l)$ is replaced with
$\sqrt{\omega_m(l)}=[l^2+m^2]^{1/4}$).
 
At the above choice of $\gamma_m(l)$, the free light-front Hamiltonian
is as follows:
\begin{equation}
\label{lfhamiltonina}
P_+=\int\, dl \,\nu_m(l)a^\dagger_l a_l,
\end{equation}
where the light-front energy of the excitation with the light-front
momentum $l$ is
\begin{equation}
\label{energy}
\nu_m(l)=\frac{\gamma^2_m(l) - l}{ \epsilon + 4l^2/M^2}.
\end{equation}
There is a qualitative difference here with the equal-time quantization.
In the equal-time quantization, the energy of the excitations is an even
function of $l$. In the light-front quantization, the excitations
with positive light-front momentum are qualitatively different
from the excitations with negative light-front momentum. The former
are the particles of mass $m$, the latter are the tachyons
of mass parameter $M$.
 
To see this, consider $\nu_m(l)$ in the limit of regularization removed,
$\epsilon\rightarrow 0, M^2\rightarrow\infty$. It has the meaning
of the momentum component along the $x^+$ direction of the excitation
whose momentum component along the
$x^-$ direction is $l$. Therefore, we expect that the product
$2l\nu_m(l)$ is independent of $l$ in this limit (because it has the
meaning
of the mass squared of the excitation). Othervise, the Lorentz invariance
is violated even in the limit of the regularization removed.
Formally, to get the expected dependence on $l$ in $\nu_m(l)$ after
the removal of the regularization at negative $l$,
we need to neglect $\epsilon$ with respect
to $4l^2/M^2$ in the denominator of (\ref{energy}). In other words, we
should take the limit $\epsilon\rightarrow 0, M\rightarrow \infty$
in such a way that $\epsilon M^2\rightarrow 0$. We also should keep
the momentum $l$ nonzero. After all these reservations,  in the limit of
the regularization removed, we have
\begin{equation}
\label{limit}
\nu_m(l)\rightarrow \theta(l)\frac{m^2}{2l} - \theta(-l)\frac{M^2}{2l},
\end{equation}
where $\theta(l)$ vanishes at negative $l$ and equals the unit at
positive $l$.
 
Note that the above limit does not hold at $l=0$:
\begin{equation}
\label{zero}
\nu_m(0)=\frac{m}{\sqrt{\epsilon}}.
\end{equation}
We see that in the regularized theory the mode of zero longitudinal
momentum
breaks Lorentz invariance. Evidently, this mode should be treated
separately.
We conjecture that this mode can be neglected at small coupling, and
causes
the phase transition at the critical coupling. We further discuss this
point
in the last section of the paper.
 
We close this section with a derivation of the propagator of the free
field in the regularized theory. The propagator is defined as
\begin{equation}
\label{propagatordef}
G_r(x)=\theta(x^+)\langle\phi(x)\phi(0)\rangle +
         \theta(-x^+)\langle\phi(0)\phi(x)\rangle.
\end{equation}
Here the time dependence of the field is obtained
from (\ref{creationannihilation}) with the replacements
\[
\exp{\big[\pm ilx^-\big]}\rightarrow\exp{\big[\pm(i\nu_m(l)x^+
+lx^-)\big]}.
\]
It is easy
to check that this time dependence is in agreement with
the equations of motion for the field implied by the regularized
Lagrangian (\ref{regularized}) at $V(\phi)=m^2\phi^2/2$.
With the above field decomposition, the propagator's Fourier
transform $\tilde{G}(k)\equiv\int \,d^2x \exp{(-ikx)}G(x)$ is
\begin{equation}
\label{fourier}
\tilde{G}(k) = \frac{1}{2i\gamma^2_m(k_-)}
\Big[\frac{1}{\nu_m(-k_-) + k_+ - i\delta} +
\frac{1}{\nu_m(k_-) - k_+ - i\delta}\Big],
\end{equation}
where $k = (k_+, k_-)$ is a two-dimensional vector.
It is easy to check that this expression does coincide with the propagator
(\ref{propagator}).
 
\section{Chang's Duality Condition}

Let us recall the derivation of Chang's duality condition
\cite{Chang:1976ek}.
If $V(\phi)=-\mu^2\phi^2/4 + g\phi^4/4$ in (\ref{lagrangian}), the small
coupling limit at negative mass squared in the Lagrangian
cannot be treated with the same field decomposition
into the creation-annihilation operators as it can be at the positive
mass squared term. This is the case because, at the negative
mass squared term, it is not possible
to remove from the free Hamiltonian the
term $a_la_{-l}+a^\dagger_la^\dagger_{-l}$ with a choice of the
real coefficient in the field's decomposition.
 
To treat the small coupling limit
at negative mass squared, one should shift the
field by a value that minimizes the potential (there are two
symmetric minima in this case).
The fluctuations of the field around the extremal value are then expanded
as in the case of the positive mass term, but with the new mass parameter
$\mu$.
The new mass parameter is defined from the second derivative of the
potential in the minimum. The Hamiltonian is then defined as a normal
ordering with respect to the creation-annihilation operators
participating in the mode expansion of the fluctuations around the
minimum.
 
Chang compares the Hamiltonian obtained in this way with the Hamiltonian
obtained for the model with the positive mass squared term by the normal
ordering with respect to the creation-annihilation operators related
to the mass parameter from the Lagrangian. To make the comparison, he
first
considers the changes induced in the Hamiltonian by the change in the
ordering
prescription caused by switching over from mass $m$ to mass $\mu$ in the
field decomposition. Next he notices that these changes transform
the Hamiltonian to the one obtained for the case with the negative mass
term
if the following duality condition is satisfied:
\begin{equation}
\label{duality}
\frac{m^2}{g} +\frac{3}{4\pi}\ln\frac{m^2}{g}=
\frac{3}{4\pi}\ln\frac{\mu^2}{g} - \frac{\mu^2}{g}.
\end{equation}
This can be interpreted as a relation between dimensionless couplings
$f\equiv g/m^2$ and $f_1\equiv g/\mu^2$. When $f$ is large, there
exists a small
$f_1$ such that the duality condition (\ref{duality}) is satisfied.
The conclusion is that the theory with a positive mass squared term and
large
coupling is equivalent to the theory with negative mass term (and nonzero
vacuum expectation of the field) and a small coupling.
 
Chang's derivation rests on the relation between normal orderings
of the products of the field operator with respect to different values of
the
mass parameter. We can repeat the reasoning of Chang for
the light-front quantization of the previous section. The only
change is in the formula relating the two ordering prescriptions.
The formula Chang uses to relate two normal orderings is
\begin{equation}
\label{normal}
N_m(e^{i\beta\phi})=(\frac{\mu^2}{m^2})^{\beta/8\pi}N_{\mu}(e^{i\beta\phi}).
\end{equation}
The factor by the normal product in the rhs, $F_{et}(\beta;\mu,m)=
(\frac{\mu^2}{m^2})^{\beta/8\pi}$
(the subscript "et" on the factor
indicates that this is the equal-time quantization), is
related
to the factor in the field decomposition as follows:
\begin{equation}
\label{eq}
F_{et}(\beta;\mu,m)=\exp{\big[\frac{1}{2}\beta\big(\Delta_{et}(\mu)-
\Delta_{et}(m)\big)\big]},
\end{equation}
where
\[
\Delta_{et}(m)=\int\,\frac{dl}{4\pi\omega_m(l)}.
\]
Here $\omega_m(l)$ is involved in the equal-time field decomposition:
\[
\phi_{et}(x^1)=\int\,\frac{dl}{\sqrt{4\pi\omega_m(l)}}\big
[a_l\exp{(-ilx^1)}+a^\dagger_l\exp{(ilx^1)}\big],
\]
where $x^1$ is the space coordinate. We can repeat all the reasoning
for the light-front quantization. The normal orderings are related in this
case by the formula
\begin{equation}
\label{normallf}
N_m(e^{i\beta\phi})=F_{lf}(\beta;\mu,m)N_{\mu}(e^{i\beta\phi}),
\end{equation}
where
\begin{equation}
\label{lf}
F_{lf}(\beta;\mu,m)=\exp{\big[\frac{1}{2}\beta\big(\Delta_{lf}(\mu)-
\Delta_{lf}(m)\big)\big]}.
\end{equation}
Here
\[
\Delta_{lf}(m)=\int\,\frac{dl}{4\pi\gamma^2_m(l)}.
\]
This is implied by the comparison of the light-front field decomposition
in (\ref{creationannihilation}), and the above field decomposition of the
equal-time quantization. We see from this comparison that the light-front
formulas for normal ordering are retained from the corresponding
equal-time
formulas with the replacement $\omega_m(l)\rightarrow \gamma^2_m(l)$.
 
We now use the explicit expression (\ref{gamma}), and check that
\[
F_{lf}(\beta;\mu,m)\rightarrow F_{et}(\beta;\mu,m)
\]
in the limit of the regularization removed. We conclude that the
relation in the light-front quantization between the normal
orderings corresponding to different masses
coincides with the one
obtained in the equal-time quantization.
 
 From this we conclude that Chang's duality relation is valid for
the light-front Hamiltonian of the previous section.
 
\section{Discussion}
 
As we have seen in the previous section, the light-front quantization of
the
theory of Eq. (\ref{regularized}) is equivalent at large coupling to
the light-front quantization of a theory obtained from Eq.
(\ref{regularized})
with a changed sign of
the mass squared term. The latter theory
should be taken at a small coupling. It has a nonzero vacuum expectation
of the field. On the other hand, at small coupling, the theory of Eq.
(\ref{regularized}) with a positive mass squared term has zero vacuum
expectation of the field. The conclusion is that somewhere on the way from
small coupling to large coupling there is a phase transition.
We recall that we reached this conclusion for the light-front quantization
of the theory whose Lagranginan is given in Eq. (\ref{regularized}).
 
At small coupling, there exist tachyons in the spectrum. The mass
parameter of the tachyons goes to infinity as the regularization
is removed. To make the above approach a computational scheme,
we should
study the decoupling of tachyons. In perturbation theory, tachyons
decouple.

A separate treatment should be given to the modes whose longitudinal
momentum is of the order $\sqrt{\epsilon}$ ($\epsilon$ is the regulator
breaking Lorentz invariance in our approach).
Perturbatively, these modes
give a finite contribution to the energy density of the vacuum in the
limit of the regularization removed (despite the fact that their
light-front
energy goes to infinity in this limit). Therefore, they are the most
viable
candidates for the role of the modes responsible for the phase transition.
The modes of small longitudinal momentum should be integrated out under
certain assumptions on their nonperturbative dynamics (the latter may
still
be out of reach for an analytical treatment). In this way, the dynamics
of conventional modes will be parameterized by an effective Hamiltonian.
It requires further study if this kind of approach will be able to
become a computational scheme.

This work was supported in part by RFBR grant no. 03-02-17047, in
part by a grant from the US Department of Energy DE-FG-87-40371, and in
part by the US National Science Foundation
NSF-PHY-007-1027. G.P. thanks Arkady Vainshtein for helpful discussions.

\end{document}